\pdfoutput=1

\documentclass[11pt]{article}

\usepackage{acl}

\usepackage{times}
\usepackage{latexsym}

\usepackage[T1]{fontenc}

\usepackage[utf8]{inputenc}

\usepackage{microtype}

\usepackage{multirow}
\usepackage{subcaption}
\usepackage{graphicx}
\usepackage{booktabs}
\usepackage{enumitem}
\graphicspath{ {./figures/} }

\title{Towards Better Semantic Understanding of Mobile Interfaces}

\author{Srinivas Sunkara$^{1*}$ \And Maria Wang$^{1*}$ \And Lijuan Liu$^{1}$ \And Gilles Baechler$^{1}$ \AND Yu-Chung Hsiao$^{1}$ \And Jindong (JD) Chen$^{1}$ \And Abhanshu Sharma$^{1}$ \And James Stout$^{2}$ \AND \\
$^1$Google Research, $^2$Google\\
}

\begin{document}
\maketitle
\begin{abstract}
Improving the accessibility and automation capabilities of mobile devices can have a significant positive impact on the daily lives of countless users. To stimulate research in this direction, we release a human-annotated dataset with approximately 500k unique annotations aimed at increasing the understanding of the functionality of UI elements. This dataset augments images and view hierarchies from RICO, a large dataset of mobile UIs, with annotations for icons based on their shapes and semantics, and associations between different elements and their corresponding text labels, resulting in a significant increase in the number of UI elements and the categories assigned to them. We also release models using image-only and multimodal inputs; we experiment with various architectures and study the benefits of using multimodal inputs on the new dataset. Our models demonstrate strong performance on an evaluation set of unseen apps, indicating their generalizability to newer screens. These models, combined with the new dataset, can enable innovative functionalities like referring to UI elements by their labels, improved coverage and better semantics for icons etc., which would go a long way in making UIs more usable for everyone.
\end{abstract}







\makeatletter
\def\blfootnote{\gdef\@thefnmark{}\@footnotetext}
\makeatother

\blfootnote{*Equal contribution, correspondence: \{srinivasksun,mariawang\}@google.com}
\section{Introduction}\label{sec:intro}

Mobile devices like phones and tablets have become ubiquitous and indispensable to carry out our daily activities. It is not an exaggeration to say that usage of mobile devices is becoming a requirement for full participation in society. 
Recent reports from the WHO and others~\cite{WHO21Blindness,Ackland17World} estimate that around 2.2 billion people across the world have some form of vision impairment, out of which 36 million people are blind. Accessibility of mobile devices is necessary for these visually impaired users to carry out their daily tasks and is an important tool for their social integration \cite{Ladner15Design}.

Accessibility of mobile apps has improved significantly over the past few years aided by developments on two main fronts: Firstly, the development of screen readers, like VoiceOver \cite{Apple20VoiceOver} on iOS and TalkBack \cite{Android20TalkBack} on Android, enable visually impaired users to control their phone in an eyes-free manner. Secondly, development tools and standards to enhance accessibility, such as the Accessibility guidelines for iOS and Android \cite{Android20AccessibilityGuidelines, Apple20AccessibilityGuidelines}, Android Accessibility Scanner \cite{Android20AccessibilityScanner}, and iOS Accessibility Inspector~\cite{Apple20AccessibilityInspector}, have helped developers identify and fix accessibility issues for applications. Among most of these utilities, the main source of accessibility data is the accessibility labels \cite{Android20Descriptions} provided by app developers. These labels are specified as attributes on a structured representation such as View Hierarchy, for the different UI elements on the screen and are available to screen readers ~\cite{Android21ViewHierarchy, Apple2018ViewHierarchy}. Despite the growth in accessibility tools, recent studies \cite{Ross2020Epidemiology, Ross2017Epidemiology, Chen2020Unblind} have found that even the most widely used apps have large gaps in accessibility. For instance, a study by \citet{Chen2020Unblind} of more than 7k apps and 279k screens revealed that around 77\% of the apps and 60\% of the screens had at least one element without explicit labels. Similarly, \citet{Ross2020Epidemiology} found that, in a population of 10k apps, 53\% of the Image Button elements were missing labels.

In this paper, we attempt to encourage further research into improving mobile device accessibility and increasing device automation by releasing an enhanced version of the RICO dataset \cite{Deka17Rico} with high-quality human annotations aimed at semantic understanding of various UI elements. Firstly, following a study by \citet{Ross2020Epidemiology} where missing labels for Image Button instances was found to be the primary accessibility barrier, we focus on creating annotations useful for identifying icons. In particular, we annotated the most frequent 77 classes of icons based on their appearance. We refer to this task as the \emph{Icon Shape} task. Secondly, we identified icon shapes which can have multiple semantic meanings and annotated each such icon with its semantic label. This task is called \emph{Icon Semantics}. Some examples of such icons can be seen in Figure~\ref{fig:semantics}. Finally, we annotate UI elements, like icons, text inputs, checkboxes etc., and associate them with their text labels. These associations can help us identify meaningful labels for the long tail of icons and UI elements not present in our schema, but having a textual label associated with them. We refer to this task as the \emph{Label Association} task. 
The main contributions of this paper are as follows:
\begin{itemize}
    \item A large scale dataset\footnote{The datasets are released at https://github.com/google-research-datasets/rico-semantics.} of human annotations for 1) Icon Shape 2) Icon Semantics and 3) selected general UI Elements (icons, form fields, radio buttons, text fields) and their associated text labels on the RICO dataset.
    \item Strong benchmark models\footnote{Benchmark models and code are released at https://https://github.com/google-research/google-research/tree/master/rico-semantics} based on state of the art models \cite{He2016Resnet, Carion2020DETR, Vaswani2017Attention} using image-only and multimodal inputs with different architectures. We present an analysis of these models evaluating the benefits of using View Hierarchy attributes and optical character recognition (OCR) along with the image pixels.
\end{itemize}
\section{Related Work}\label{sec:related-work}
\subsection{Datasets}
Large scale datasets like ImageNet \cite{Deng2009ImageNet} played a crucial part in the development of Deep Learning models~\cite{krizhevsky2012Alexnet, He2016Resnet} for Image Understanding. Similarly, the release of the RICO dataset \cite{Deka17Rico} enabled data driven modeling for understanding user interfaces of mobile apps. RICO is, to the best of our knowledge, the largest public repository of mobile app data, containing 72k UI screenshots and their View Hierarchies from 9.7k Android apps spanning 27 categories. Apart from RICO, other datasets include ERICA \cite{Deka2016Erica} with sequences of user interactions with mobile UIs and LabelDROID \cite{Chen2020Unblind} which contains 13.1k mobile UI screenshots and View Hierarchies.

\looseness=-1
There have been a few efforts to provide additional annotations on RICO. SWIRE \cite{Huang2019Swire} and VINS \cite{Bunian2021VINS} added annotations for UI retrieval, Enrico \cite{Leiva2020Enrico} added annotations for 20 design topics. \citet{Liu2018LearningDesign} automatically generated semantic annotations for UI elements using a convolutional neural network trained on a subset of the data. Recently \citet{li2022learningtodenoise} released UI element labels on view hierarchy boxes including identifying boxes which do not match to any elements in the UI. Even though some of these works are similar in spirit to the dataset presented in this paper, there are two major differences: 1) The icon and UI element labels are \emph{inferred} on the boxes extracted from the View Hierarchy, whereas, in our work, we add \emph{human annotated} bounding boxes directly on the image. Due to noise in the view hierarchies like missing and misaligned bounding boxes for UI elements \cite{li2020RicoSCA, li2022learningtodenoise}, we observe that human annotation increases the number of icons labelled by 47\%. 2) The semantic icon labels in \citet{Liu2018LearningDesign} conflate appearance and functionality. For example ``close'' and ``delete,'' ``undo'' and ``back,'' ``add'' and ``expand'' are mapped to the same class, even though they represent different functionalities. The \emph{Icon Semantics} annotations in our dataset specifically try to distinguish between icons with the same appearance but differences in functionality.

\begin{figure*}[t!]
    \centering
    \begin{subfigure}[t]{0.7\textwidth}
        \centering
        \includegraphics[width=\textwidth]{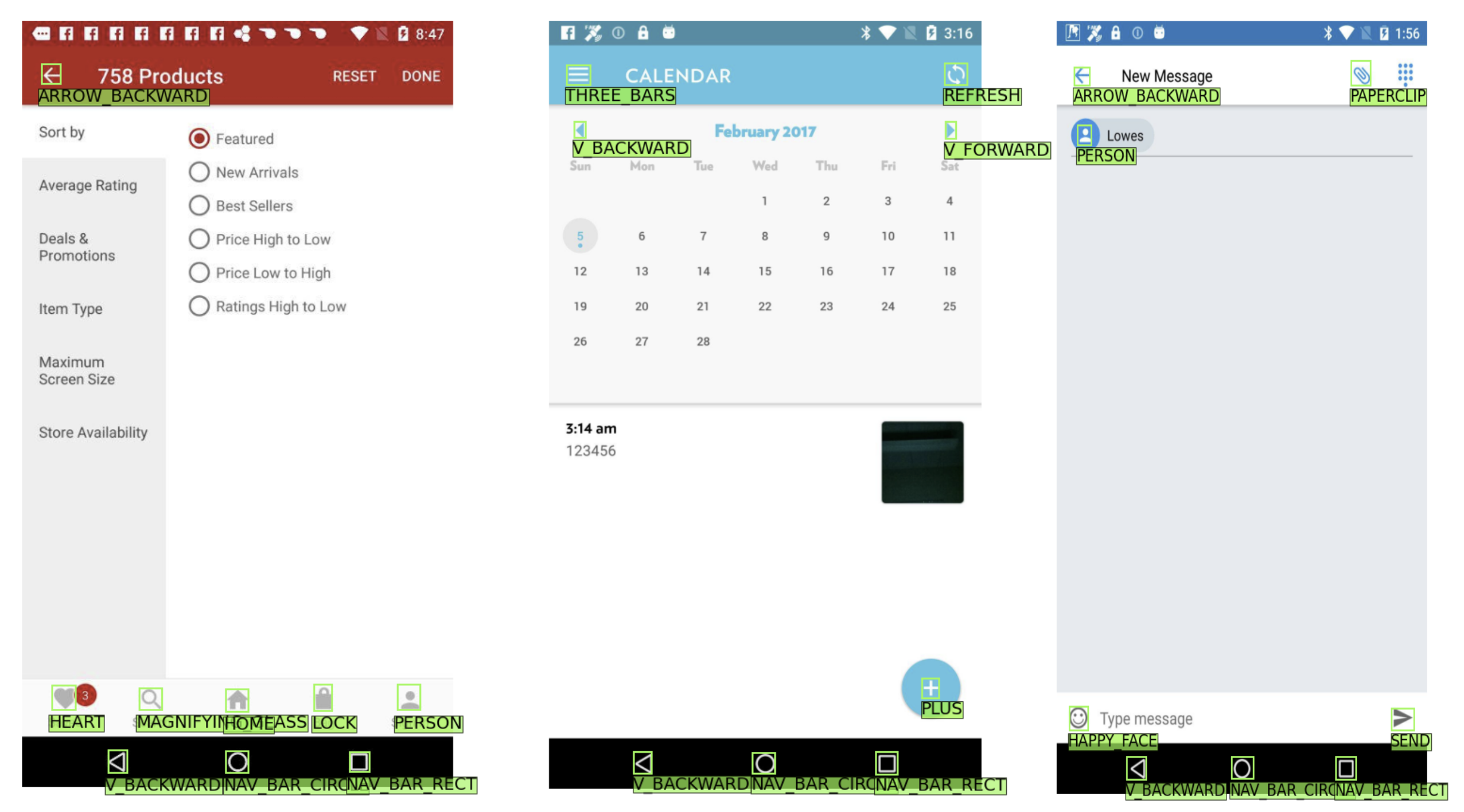}
        \caption{Examples of 76 \emph{icon shape} annotations. We include classes that reflect the social aspects of app usage, e.g., "person" for profile and community, share via popular apps such as Facebook and Twitter, etc.}
        \label{fig:shape-examples}
    \end{subfigure}
    \begin{subfigure}[t]{0.7\textwidth}
        \centering
        \includegraphics[width=\textwidth]{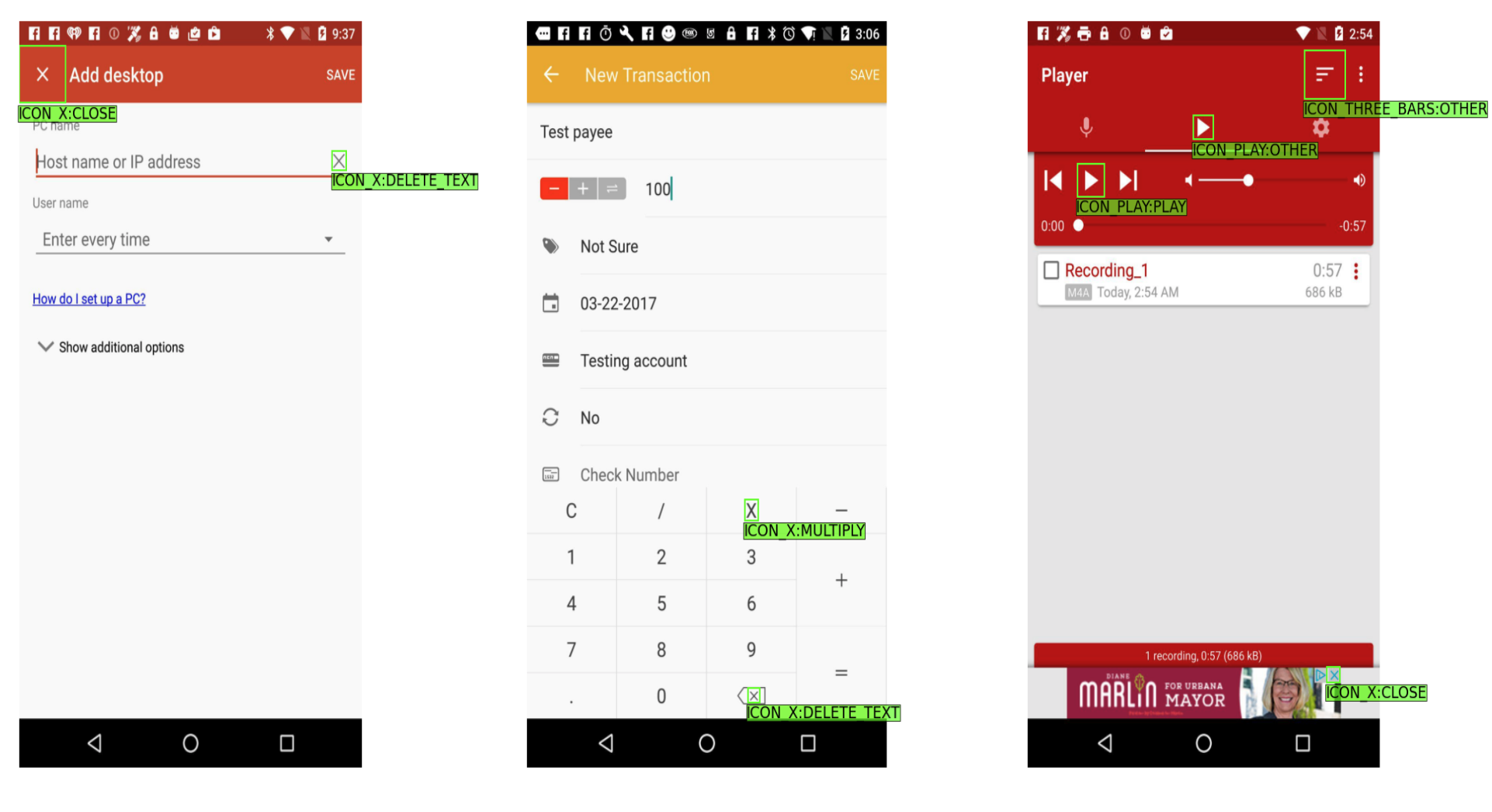}
        \captionsetup{belowskip=-5pt}
        \caption{Examples of 38 \emph{icon semantics} annotations, in the format of <shape>:<semantics>. Note that a single shape may represent multiple semantics, depending on the context. E.g., "X" shape may mean "close", "delete text", or "multiply". We use an umbrella semantic class "OTHER" to cover the semantics not covered in our proposed set of classes}
        \label{fig:semantics}
    \end{subfigure}
    \captionsetup{belowskip=-10pt}
    \caption{In this paper, we annotated the RICO dataset with both \textbf{icon shapes} and their \textbf{semantics}, to encourage further research on app automation and accessibility. The existing icon annotations from~\citet{Liu2018LearningDesign} were algorithmically generated with 10\% of them verified. However, we observed $\sim 32$\% were missing labels compared to our full human annotations. We release our annotations in the hope to contribute back to the community.}
    \label{fig:shape-and-semantics}
\end{figure*}

\subsection{Models}
Pixel based methods for UI understanding have been studied for a long time. They have been used for a variety of applications like GUI testing \cite{Yeh2009Sikuli, Chang2015GUI}, identifying similar products across screens \cite{Bell2015Learning}, finding similar designs and UIs \cite{Behrang2018GUIFetch, Bunian2021VINS}, detecting issues in UIs \cite{Liu2020Discovering}, generating accessibility metadata \cite{Zhang2021Screen} and generating descriptions of elements \cite{Chen2020Unblind}. A recent study by \citet{Chen21Object} compares traditional image processing methods and Deep Learning methods to identify different UI elements. The image-only baseline models studied in this paper are based on Object Detection methods presented in \citet{Chen21Object}, \citet{Zhang2021Screen}, \citet{Chen2020Unblind}, and \citet{Carion2020DETR}.

Extending to other modalities beyond pixels,
\citet{Banovic2021Waken} use video tutorials to understand UIs and annotate them with additional information. \citet{Li2021Screen} use only the screen information for identifying embeddings of UI elements. \citet{Hurst2010Auotmatically} use both the screen and accessibility API information to identify interaction targets in UIs and \citet{Chang2011Associating} use similar inputs to detect and identify certain UI elements and \citet{Tam2018Deep} use it for identifying similar UI designs. Multimodal inputs have also been used for understanding screen contents like generating element descriptions~\cite{Li2020Widget}, training UI embeddings for multiple downstream tasks~\cite{He2021ActionBert, Bai2021UIBert} and, denoising data, predicting bounding box types~\cite{li2022learningtodenoise}. 
\section{Datasets, taxonomy and annotation}\label{sec:datasets}
To enable accessible hands-free experience for mobile users, it is necessary for the system to understand the functionality of the different screen UI elements. For learning data driven models to enable these functionalities, we use the RICO dataset \cite{Deka17Rico}. RICO spans >9K apps and >72K UIs, each with a screenshot and information regarding the structure of the UI in the form of a View Hierarchy (VH). Besides the bounding boxes of the different UI elements, the VH contains useful attributes like the \emph{content description} and \emph{resource id} which provide information regarding the functionality of the different UI elements.  However, \citet{li2020RicoSCA} found that only 35\% of the unique screens in RICO contain a matching View Hierarchy and screenshot. In the next few sections, we describe how we overcome the mismtach issue and describe the different type of annotations.

\subsection{Icon Shape}\label{sec:icon-shape}
As the study by \citet{Ross2020Epidemiology} found that one of the main accessibility barriers on mobile devices are missing labels for ImageButton elements, in the \emph{icon shape} and \emph{semantics} tasks we focus on creating annotations useful for identifying icons. \citet{Liu2018LearningDesign} attempted to provide semantic labels for the icons in the RICO dataset by identifying different concepts represented by the ImageButton elements on a subset of the data. Using these concepts and models trained on a subset of the data they identified semantic labels over the entire dataset. They inferred labels for more than 100 icons types, 25 UI element types and 197 text button types. 
However, due to the presence of view hierarchies with bounding boxes missing and misaligned for UI elements \cite{li2020RicoSCA, li2022learningtodenoise}, these labels miss several UI elements. By comparing with manually labeled data, we found that the annotations in~\citet{Liu2018LearningDesign} did not identify around 32\% of all icon instances captured by manual labeling.
For improving the coverage of icons, in our dataset we asked raters to manually annotate all the bounding boxes. We created a schema of the 77 most commonly used icon classes, reusing many of the classes identified in \citet{Liu2018LearningDesign}. Examples of these icon classes and images are shown in Figure~\ref{fig:shape-examples}.

\subsection{Icon Semantics}\label{sec:icon-semantics}
 To support voice driven usage of mobile devices, we identify icons not only based on their shape and appearance but also functionality and semantic meaning. For example, an ``X'' shaped icon can mean ``close,'' ``remove an option/entry,'' ``delete/clear text,'' or ``multiply.'' One way to enable users to refer to the various semantics is to map the multiple semantics to the same class in the \emph{Icon Shape} schema. This approach has two limitations: 1) Mapping each icon shape to multiple semantics can lead to confusion for applications like Screen Readers. 2) We noticed that there are many instances of icons with different semantics but same shape occurring in the same screen. In particular 11\% of all images with \emph{ICON\_PLAY}, 4.9\% with \emph{ICON\_X} and 4.8\% with \emph{ICON\_CHAT} have icons with multiple semantics.

We identified a list of common icons which have more than one semantic meaning associated with them by the following steps: 
1) Manually inspect a variety of the icon annotations and list the functionality of each icon instance,
2) Use the plurality of words matching to the same icon shape in the RICO icon labels in \cite{Liu2018LearningDesign} as an indicator of multiple semantics, and 
3) Recognize confusing app logos. For example, the logo for WhatsApp contains a Phone icon but it is most natural for a user to say, ``open WhatsApp''.

After these steps, we arrived at a list of 12 shape icons which were further classified into 38 semantic shapes. For icons with semantic meanings not covered by our schema, we assign the semantic type \emph{OTHER} as the default label. Out of the 101,625 annotated icons 15,640 (around 15\%) are labeled as \emph{OTHER}. We observed that it is difficult to cover all of the tail semantics classes with a schema. Thus, we also obtained annotations for the text labels associated with UI elements, described in the section below. 
\subsection{Label Association}
\begin{figure*}[t]
    \centering
        \centering
        \includegraphics[width=0.65\textwidth]{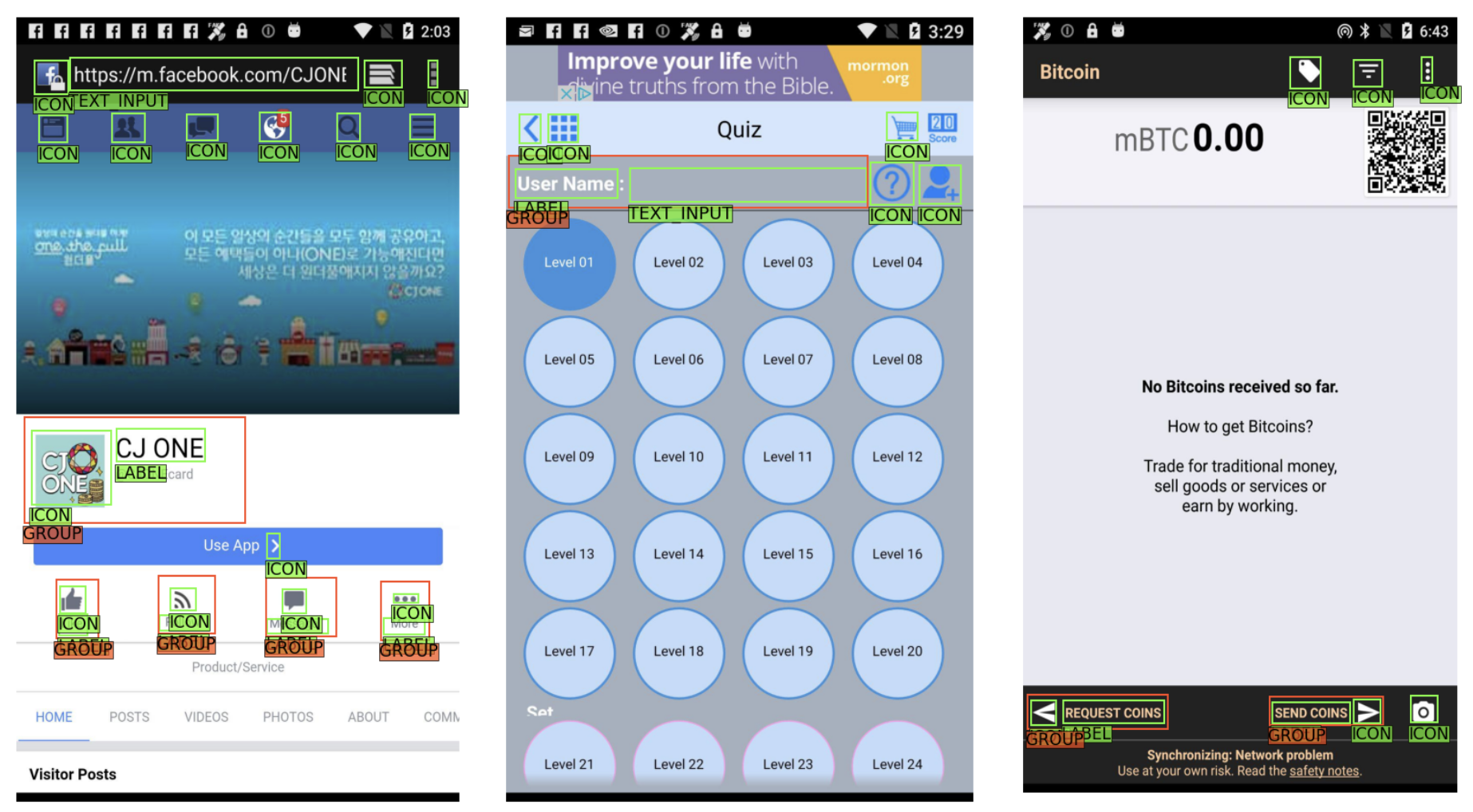}
\captionsetup{belowskip=-10pt}
\caption{When an associated text label appears with a UI element, it is natural to refer to the icon using the text directly. 24.6\% of icons, form fields, check boxes and radio buttons have an associated text label. \emph{Label Association} examples are provided in the figure in red. However, rule-based approaches using view hierarchies or OCR can only achieve 40\% and 29.5\% accuracies respectively. These annotations enable us to identify the elements to interact with for voice commands like ``Enter User Name as test user'' or ``Request Coins''.}
\label{fig:grouping-examples}
\end{figure*}

Many UI elements have an associated text labels that best describes UI elements. Our data analysis showed that 24.6\% of icons, form fields, check boxes and radio buttons have an associated label. However, we found that these labels are commonly neither syntactically associated within view hierarchies nor visually aligned in screenshot pixels. First, we attempted to identify labels associated with UI elements by using heuristics relying on the View Hierarchy, like searching the siblings and parent node’s siblings for axis aligned text elements for each of the UI elements. We found that only~40\% of UI elements with labels could be correctly associated using such heuristics with a significant number of false positive associations. Next, we attempted to match UI element bounding boxes with line boxes detected by OCR. We matched each UI element bounding box with the OCR text box nearest to the top left corner with a maximum distance threshold. This method achieved~29.5\% accuracy. These empirical studies indicated that, like many machine perception tasks, this {\emph Label Association} task may in fact be non-trivial despite appearing straightforward to humans. Some examples can be seen in figure~\ref{fig:grouping-examples}. We believe this novel task can help address the limitation of annotations with a fixed set of classes by making use of the text label information present in the UI. 
\begin{figure}[t]
\includegraphics[width=0.48\textwidth]{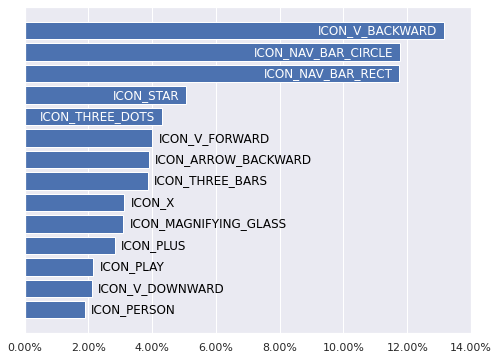}
\caption{Distribution of the top 14 \emph{icon shape} classes. These classes account for 72\% of the total icons covered by the \emph{icon shape} labels.}
\label{fig:icon-shape-distr-bar}
\end{figure}
\hfill
\begin{figure}[t]
\includegraphics[width=0.48\textwidth]{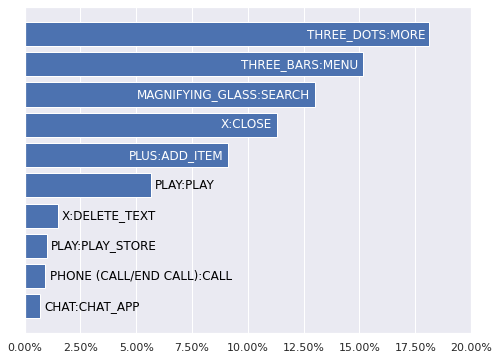}
\caption{Distribution of the top 10 \emph{icon semantics} classes. These classes account for 76\% of the total icons covered by the \emph{icon semantics} labels.}
\label{fig:icon-sem-distr-bar}
\end{figure}

\subsection{Annotation Procedure}
For the three tasks described above, we follow the annotation procedure below to collect annotated data. We used a team of 40 trained human raters with single replication to annotate the screenshots. The raters were initially provided with example image patches for each icon or UI element type similar to Figure~\ref{fig:shape-examples} and then followed the annotation procedure below:
\begin{itemize}
    \item Round 1: Each rater is presented with a screenshot and is asked to draw a bounding box around every icon and UI element on the screen. For each icon box, they can specify an icon shape class if it is included in the schema and otherwise they classify it as a general icon class. Once an icon shape class is identified, if there are multiple semantics associated with that shape, the raters choose one semantic class among the different options for that class. These options also include the \emph{OTHER} class to capture semantics not in the icon semantics schema. For the label association task, we ask raters to identify the icons, form fields, radio buttons, and check boxes first, followed by the text labels, by drawing their bounding boxes and assigning the respective classes. Then the raters group the text labels with their associated UI element if it exists.
    \item Round 2: To improve the annotation quality, we send the datasets for a second round of cleaning where the raters can adjust the bounding box or the classes assigned.
    \item Round 3: After round 2, if we still find some classes are poorly labeled by manual inspection, we use trained models to identify potential incorrect labels. We train Object Detection models on the entire dataset, and use the model to predict labels for the train, validation and test sets. If there are any differences between the model predictions and the human annotations, we identify these instances as potentially error-prone and send these images back to the raters to re-annotate them.
\end{itemize}
Using the above procedure, we annotated all of the screenshots in RICO with icon shape, semantics and label association classes. The distribution of top labels for the icon tasks can be seen in figures~\ref{fig:icon-shape-distr-bar},~\ref{fig:icon-sem-distr-bar}. The entire taxonomy and exact counts are in Appendix~\ref{sec:appendix-data}.
\section{Baseline Models}
We conducted several experiments to investigate the effectiveness of various deep learning approaches for solving the tasks presented in Section~\ref{sec:datasets}. The overall goal of these experiments is to: 1) provide good baseline models to be used for image-only and image+VH settings 2) study the effect of using multi-modal inputs v/s only the screenshot for these tasks. 
\subsection{Problem Setup}
For the three tasks, we distinguish two different approaches: 1) the Object Detection (OD) approach using only the image and, 2) the bounding box classification (BB-CLS) approach using image, OCR and view hierarchy. We describe these two approaches in the following sections. We split the data into 80\% train, 10\% validation and 10\% test by package name to avoid data leakage and use the same split for all experiments.
\subsubsection{Object Detection (OD)}
In this setup, the models take the screenshot as input and output bounding box, class label and score for each object found. We used object detection models based on the widely used and better performing~\cite{Chen21Object} Faster R-CNN \cite{ren2015frcnn} and Centernet \cite{zhou2019centernet} architectures. We train these models with various backbones \cite{szegedy2017inception,He2016Resnet,lin2017fpn,Newell2016hourglass} and report results for the best performing ones.
Finally, we experimented with Object Detection models based on  Transformer \cite{Vaswani2017Attention} architecture like DETR \cite{Carion2020DETR, zhu2020deformable} to verify the hypothesis that for the \emph{Icon Semantics} and the \emph{Label Association} tasks, the models need more information from their context compared to the \emph{Icon Shape} task. We use standard Object Detection metrics like mAP@0.5IOU to compare the model performance.
\subsubsection{Bounding box classification (BB-CLS)}
For this setup, we train models to classify bounding boxes extracted from the view hierarchy (VH) and assign them labels among the candidate classes. The groundtruth (GT) set contains boxes that have been created and labeled by crowd workers.
These two sets of boxes are greedily matched as follows. 1) Each GT box is matched to at most one VH box and vice versa. 2) For every GT box, we find the VH box with maximum Intersection over Union (IoU). Only the matches for which the IoU value is greater than the threshold of~\emph{0.5} are kept. 3) Once a VH box is matched with a GT, it is not considered for future matches. After this matching procedure is complete, we assign a background class to the unmatched VH boxes. We use a Transformer \cite{Vaswani2017Attention} based network as described in UIBert \cite{Bai2021UIBert} for the \emph{Icon Shape}, \emph{Semantics} tasks framed as a classification problem and compare models based on Macro F-1 score. 

For the \emph{Label Association} task, we compute an embedding of each UI element following~\citet{Bai2021UIBert} and perform clustering on projected embeddings to identify UI elements which belong to a group. We use F-1 score of the associated elements as the metric of comparison. For every set of elements predicted to be a group, it is considered a True Positive if the same group is present in the groundtruth, and considered a False Positive if it is not. All groundtruth groups which are not in predictions count as False Negatives and the F-1 score is computed based on these counts.

Compared to OD models the BB-CLS models have an advantage as they do not need to predict the bounding boxes for UI elements. To enable a comparison between the two approaches, we add the unmatched GT boxes as inputs by setting all the other VH attributes to be empty values. To use this in a real-world setting, this procedure assumes the existence of a good VH without missing boxes. We study BB-CLS models here as it helps us validate the potential benefit of using VH attributes from the UI elements on the screen and consequently can motivate improving the view hierarchies for various apps and web-sites.

\subsection{Model Configurations and Training Details}
For the Centernet model, we used the hourglass-104 backbone~\cite{zhou2019centernet} with an input size of 1024$\times$1024. For the DETR model~\cite{Carion2020DETR}, each image is proportionally resized and padded to the shape 1280$\times$1280. We use a ResNet-50~\cite{He2016Resnet} pretrained on ImageNet~\cite{Deng2009ImageNet} as the backbone with frozen batch normalization layers for training stability. We add position embedding and object queries to each layer. The DETR models are trained on cloud TPUs with a batch size of 256 and reduce learning rate from $1\times10^{-4}$ to $1\times10^{-5}$ after 120k steps. 

For classification and clustering models based on UIBert \cite{Bai2021UIBert}, we use an EfficientNet-B0 \cite{tan2019efficientnet} model for encoding the image patches and use ALBERT text encoder \cite{lan2019albert} for encoding OCR and VH attributes. We use a Transformer layer with 6 layers, 16 heads and a intermediate size of 512. We train these models on cloud TPUs with a batch size of 128 using the Adam optimizer \cite{kingma2014adam} with a warmup over 10k steps and reduce learning rate from $1\times10^{-4}$ by a factor of 3 for every 50k steps . All the models were implemented using Tensorflow \cite{tensorflow2015-whitepaper} and converged within two days.
\begin{table*}[t]
\begin{subtable}[h]{0.48\textwidth}
    \centering
    \begin{tabular}{l c c} 
    \toprule
    Model Type & mAP & mAP@0.5IOU  \\
    \hline
    Faster R-CNN & 34.60 & 70.24 \\
    CenterNet Hourglass & 37.50 & 72.50 \\
    DETR & \textbf{39.28} & \textbf{77.94}\\
    \hline
    \end{tabular}
    \caption{Object Detection}
\end{subtable}
\hfill
\begin{subtable}[h]{0.48\textwidth}
    \centering
    \begin{tabular}{l c c}
    \toprule
    Inputs & F-1 score  & 95\% CI\\
    \hline
    Image + OCR + VH & \textbf{83.38} & [83.23 - 83.53] \\
    Image + OCR & 82.75 & [82.54 - 82.95] \\
    Image only & 82.08 & [81.71 - 82.44] \\
    \hline
    \end{tabular}
    \caption{Bounding box Classification}
\end{subtable}
\caption{Baseline model performance on \emph{Icon Shape} task. For the object detection models, DETR outperforms CenterNet and Faster R-CNN architectures. For bounding box classification, models using the image OCR and view hierarchy outperform ones not using all modalities.}
\label{table:shape_results}
\vspace{-1mm}
\end{table*}

\begin{table*}[t!]
\begin{subtable}[h]{0.48\textwidth}
    \centering
    \begin{tabular}{l c c} 
    \toprule
    Model Type & mAP & mAP@0.5IOU  \\
    \hline
    Faster R-CNN & 25.33 & 53.59 \\
    CenterNet Hourglass & 25.70 & 54.00 \\
    DETR & \textbf{26.69} & \textbf{55.74} \\
    \hline
    \end{tabular}
    \caption{Object Detection}
\end{subtable}
\hfill
\begin{subtable}[h]{0.48\textwidth}
    \centering
    \begin{tabular}{l c c}
    \toprule
    Inputs & F-1 score & 95\% CI \\
    \hline
    Image + OCR + VH & \textbf{67.16} & [66.74 - 67.58]\\
    Image + OCR & 64.52 & [63.98 - 65.05] \\
    Image only & 63.66 & [63.15 - 64.16] \\
    \hline
    \end{tabular}
    \caption{Bounding box Classification}
\end{subtable}
\caption{Baseline model performance on \emph{Icon Semantics} task. Similar to the For the \emph{Icon Shape} task DETR is the best performing object detection model and models using image, OCR and view hierarchy perform the best. However, there is no statistically significant improvement between image only and image + OCR models.}
\label{table:semantics_results}
\vspace{-1mm}
\end{table*}

\begin{table*}[t!]
\begin{subtable}[h]{0.48\textwidth}
    \centering
    \begin{tabular}{l c c} 
    \toprule
    Model Type & mAP & F-1 score  \\
    \hline
    Faster R-CNN & 36.90 & 75.75 \\
    CenterNet Hourglass & 38.00 & 75.65 \\
    DETR & \textbf{40.71} & \textbf{79.17} \\
    \hline
    \end{tabular}
    \caption{Object Detection}
\end{subtable}
\hfill
\begin{subtable}[h]{0.48\textwidth}
    \centering
    \begin{tabular}{l c c}
    \toprule
    Inputs & F-1 score & 95\% CI \\
    \hline
    Image + OCR + VH & \textbf{87.23} & [86.40 - 88.05] \\
    Image + OCR & 85.29 & [84.78 - 85.79] \\
    Image only & 84.33 & [83.30 - 85.36] \\
    \hline
    \end{tabular}
    \caption{Bounding box Classification}
\end{subtable}
\caption{Baseline model performance on \emph{Label Association} task. The model performance characteristics are very similar to those observed for the \emph{Icon Semantics} task.}
\label{table:grouping_results}
\vspace{-1mm} 
\end{table*}
\subsection{Results and Analysis}
\label{section:results}
In this section, we report our model performance for each problem setup. For all model variants, we choose the model with the best performance on the validation set and report the numbers on the test set. Overall, we observe that for \emph{BB-CLS} models using Image + VH + OCR perform better than models using Image + OCR and Image only. For \emph{OD} models, we report the results from the best performing Faster R-CNN model, the best performing CenterNet model, and DETR, the overall best model for each task. We report both these results as it allows us to compare the performance of CNN based architectures with Transformer based ones. Also CenterNet models enable fast inference \cite{duan2019centernet} on mobile phones compared to DETR models, making the baseline models directly usable on mobile phones. Results for all the different architectures we experimented with can be found in the Appendix~\ref{sec:appendix-data}. For \emph{BB-CLS} models, we estimate the 95\% Confidence Intervals based on 5 model runs with the same configuration.
\subsubsection{Icon Shape and Semantics}
Among \emph{OD} models, we observe that models based on DETR \cite{Carion2020DETR} which uses Transformers + Convolutions outperform CNN-based object detection models. DETR models achieve an mAP@0.5IOU of 77.94\% on the test set vs 72.50\% for CenterNet models on \emph{Icon Shape} task and achieve an mAP@0.5IOU of 55.74\% vs 54\% for CenterNet models for \emph{Icon Semantics} task. The performance of all of the models is weaker for \emph{Icon Semantics} vs \emph{Icon Shape} task. We believe this is due to \emph{Icon Semantics} being a harder task as objects of similar shape can belong to different classes based on the rest of the screenshot or other assumptions. Additionally, since semantic labels are a sub-classification of shape labels, this dataset has fewer labeled examples per class. 

For the \emph{BB-CLS} models, we observe that models which take the VH as input outperform models without VH for both tasks. Models with Image + VH + OCR outperform models with Image + OCR by 0.63\% and 2.6\% for the \emph{Shape} and \emph{Semantics} tasks respectively. We believe this could be a result of information regarding the elements being present in VH attributes like content description. However, adding OCR does not seem to improve model performance significantly over using only the image as input. Detailed results for these tasks can be found in Tables~\ref{table:shape_results},~\ref{table:semantics_results}.
\subsubsection{Label Association}
For the \emph{Label Association} task, DETR-based models also outperform CenterNet models in terms of F-1 score of 79.17\% vs 75.65\%. For BB-CLS models, we observe that models using VH attributes outperform models not using VH with an F-1 score of 87.23\% vs 85.29\%. Detailed results can be found in Table~\ref{table:grouping_results}. We observe a significant gap in the F1-score achieved by BB-CLS models v/s OD models.
\section{Applications and Future Work}
\label{sec:applications}
This dataset and models built on it to predict the icon and text association labels can be used to improve the label coverage for various accessibility applications like VoiceOver \cite{Apple20VoiceOver}, TalkBack \cite{Android20TalkBack}, Voice Access \cite{Android22VoiceAccess} etc. In addition, these labels can be used to improve the accessibility labels for various other platforms like web browsers and desktop applications. The models can also be used to automatically suggest accessibility labels for UI elements based on their appearance in various developer platforms~\cite{XCode2022iOS,AndroidStudio2022} so that developers can improve the accessibility of their apps easily. Along with improving accessibility, we believe this dataset is a step towards enabling new features like voice control, screen summarization and others.\par
The baseline models presented in this paper can be improved in a number of ways like training multitask models for a single model to output the different labels, bridge the gap between the models which use only the image and models which use view hierarchy, improve the ability of multimodal models to handle missing view hierarchy elements and their attributes. Other directions of research include inferring labels for the long tail icon classes using the \emph{ICON} class annotations from the \emph{label association} task, inferring semantic labels for general UI elements. This supervised data can also be used to improve the performance of self-supervised methods for UI elements like ActionBert \cite{He2021ActionBert}, UIBert \cite{Bai2021UIBert} etc.
\section{Conclusions}\label{sec:conclusions}
In this paper, we presented an enhanced version of the RICO dataset with three new sets of annotations aimed at improving the semantic understanding of mobile screens, namely \emph{icon shape}, \emph{icon semantics}, and \emph{label association}. We outlined the benefits of human annotated data over automatically labeled data and released strong baseline models using image and view hierarchy for each of these tasks. Our dataset, benchmark models and experiments lay the groundwork for future research on building better models for semantic understanding of UIs. We observe that using pre-trained models and view hierarchy attributes is a promising direction for improving these models. These models can be combined with other techniques like heuristic rules to infer the multitude of labels useful for driving improvements in accessibility and automation of mobile devices.

\bibliographystyle{acl_natbib}
\bibliography{bibliography}

\begin{thebibliography}{57}
\expandafter\ifx\csname natexlab\endcsname\relax\def\natexlab#1{#1}\fi

\bibitem[{Abadi et~al.(2015)Abadi, Agarwal, Barham, Brevdo, Chen, Citro,
  Corrado, Davis, Dean, Devin, Ghemawat, Goodfellow, Harp, Irving, Isard, Jia,
  Jozefowicz, Kaiser, Kudlur, Levenberg, Man\'{e}, Monga, Moore, Murray, Olah,
  Schuster, Shlens, Steiner, Sutskever, Talwar, Tucker, Vanhoucke, Vasudevan,
  Vi\'{e}gas, Vinyals, Warden, Wattenberg, Wicke, Yu, and
  Zheng}]{tensorflow2015-whitepaper}
Mart\'{i}n Abadi, Ashish Agarwal, Paul Barham, Eugene Brevdo, Zhifeng Chen,
  Craig Citro, Greg~S. Corrado, Andy Davis, Jeffrey Dean, Matthieu Devin,
  Sanjay Ghemawat, Ian Goodfellow, Andrew Harp, Geoffrey Irving, Michael Isard,
  Yangqing Jia, Rafal Jozefowicz, Lukasz Kaiser, Manjunath Kudlur, Josh
  Levenberg, Dandelion Man\'{e}, Rajat Monga, Sherry Moore, Derek Murray, Chris
  Olah, Mike Schuster, Jonathon Shlens, Benoit Steiner, Ilya Sutskever, Kunal
  Talwar, Paul Tucker, Vincent Vanhoucke, Vijay Vasudevan, Fernanda Vi\'{e}gas,
  Oriol Vinyals, Pete Warden, Martin Wattenberg, Martin Wicke, Yuan Yu, and
  Xiaoqiang Zheng. 2015.
\newblock \href {https://www.tensorflow.org/} {{TensorFlow}: Large-scale
  machine learning on heterogeneous systems}.
\newblock Software available from tensorflow.org.

\bibitem[{Accessibility(2021{\natexlab{a}})}]{Android20AccessibilityGuidelines}
Android Accessibility. 2021{\natexlab{a}}.
\newblock \href
  {https://developer.android.com/guide/topics/ui/accessibility/apps}
  {Accessibility guidelines}.

\bibitem[{Accessibility(2021{\natexlab{b}})}]{Android20AccessibilityScanner}
Android Accessibility. 2021{\natexlab{b}}.
\newblock \href
  {https://play.google.com/store/apps/details?id=com.google.android.apps.accessibility.auditor}
  {Accessibility scanner}.

\bibitem[{Accessibility(2021{\natexlab{c}})}]{Android21ViewHierarchy}
Android Accessibility. 2021{\natexlab{c}}.
\newblock \href
  {https://developer.android.com/reference/android/view/accessibility/AccessibilityNodeProvider}
  {Accessibilitynodeprovider}.

\bibitem[{Accessibility(2021{\natexlab{d}})}]{Android20Descriptions}
Android Accessibility. 2021{\natexlab{d}}.
\newblock \href
  {https://developer.android.com/guide/topics/ui/accessibility/apps#describe-ui-element}
  {Make apps more accessible}.

\bibitem[{Accessibility(2021{\natexlab{e}})}]{Android20TalkBack}
Android Accessibility. 2021{\natexlab{e}}.
\newblock \href
  {https://support.google.com/accessibility/android/answer/6283677?hl=en}
  {Talkback}.

\bibitem[{Accessibility(2022)}]{Android22VoiceAccess}
Android Accessibility. 2022.
\newblock \href
  {https://support.google.com/accessibility/android/topic/6151842?hl=en&ref_topic=9079844}
  {Voice access on android}.

\bibitem[{AndroidStudio(2021)}]{AndroidStudio2022}
AndroidStudio. 2021.
\newblock \href {https://developer.android.com/studio} {Androidstudio}.

\bibitem[{Apple(2018)}]{Apple2018ViewHierarchy}
Apple. 2018.
\newblock \href
  {https://developer.apple.com/library/archive/documentation/General/Conceptual/Devpedia-CocoaApp/View\%20Hierarchy.html}
  {View hierarchy}.

\bibitem[{Apple(2021{\natexlab{a}})}]{Apple20AccessibilityInspector}
Apple. 2021{\natexlab{a}}.
\newblock \href
  {https://developer.apple.com/library/archive/technotes/TestingAccessibilityOfiOSApps/TestAccessibilityiniOSSimulatorwithAccessibilityInspector/TestAccessibilityiniOSSimulatorwithAccessibilityInspector.html}
  {Accessibility inspector}.

\bibitem[{Apple(2021{\natexlab{b}})}]{Apple20AccessibilityGuidelines}
Apple. 2021{\natexlab{b}}.
\newblock \href
  {https://developer.apple.com/design/human-interface-guidelines/accessibility/overview/introduction/}
  {Accessibility overview}.

\bibitem[{Apple(2021{\natexlab{c}})}]{Apple20VoiceOver}
Apple. 2021{\natexlab{c}}.
\newblock \href
  {https://support.apple.com/guide/iphone/turn-on-and-practice-voiceover-iph3e2e415f/ios}
  {Voiceover}.

\bibitem[{Bai et~al.(2021)Bai, Zang, Xu, Sunkara, Rastogi, Chen, and Agüera~y
  Arcas}]{Bai2021UIBert}
Chongyang Bai, Xiaoxue Zang, Ying Xu, Srinivas Sunkara, Abhinav Rastogi,
  Jindong Chen, and Blaise Agüera~y Arcas. 2021.
\newblock \href {https://doi.org/10.24963/ijcai.2021/235} {Uibert: Learning
  generic multimodal representations for ui understanding}.
\newblock In \emph{Proceedings of the Thirtieth International Joint Conference
  on Artificial Intelligence, {IJCAI-21}}, pages 1705--1712. International
  Joint Conferences on Artificial Intelligence Organization.
\newblock Main Track.

\bibitem[{Banovic et~al.(2012)Banovic, Grossman, Matejka, and
  Fitzmaurice}]{Banovic2021Waken}
Nikola Banovic, Tovi Grossman, Justin Matejka, and George Fitzmaurice. 2012.
\newblock \href {https://doi.org/10.1145/2380116.2380129} {Waken: Reverse
  engineering usage information and interface structure from software videos}.
\newblock In \emph{Proceedings of the 25th Annual ACM Symposium on User
  Interface Software and Technology}, UIST '12, page 83–92, New York, NY,
  USA. Association for Computing Machinery.

\bibitem[{Behrang et~al.(2018)Behrang, Reiss, and Orso}]{Behrang2018GUIFetch}
Farnaz Behrang, Steven~P. Reiss, and Alessandro Orso. 2018.
\newblock \href {https://doi.org/10.1145/3197231.3197244} {Guifetch: Supporting
  app design and development through gui search}.
\newblock In \emph{Proceedings of the 5th International Conference on Mobile
  Software Engineering and Systems}, MOBILESoft '18, page 236–246, New York,
  NY, USA. Association for Computing Machinery.

\bibitem[{Bell and Bala(2015)}]{Bell2015Learning}
Sean Bell and Kavita Bala. 2015.
\newblock \href {https://doi.org/10.1145/2766959} {Learning visual similarity
  for product design with convolutional neural networks}.
\newblock \emph{ACM Trans. Graph.}, 34(4).

\bibitem[{Bunian et~al.(2021)Bunian, Li, Jemmali, Harteveld, Fu, and Seif
  El-Nasr}]{Bunian2021VINS}
Sara Bunian, Kai Li, Chaima Jemmali, Casper Harteveld, Yun Fu, and Magy~Seif
  Seif El-Nasr. 2021.
\newblock \href {https://doi.org/10.1145/3411764.3445762} {Vins: Visual search
  for mobile user interface design}.
\newblock In \emph{Proceedings of the 2021 CHI Conference on Human Factors in
  Computing Systems}, CHI '21, New York, NY, USA. Association for Computing
  Machinery.

\bibitem[{Carion et~al.(2020)Carion, Massa, Synnaeve, Usunier, Kirillov, and
  Zagoruyko}]{Carion2020DETR}
Nicolas Carion, Francisco Massa, Gabriel Synnaeve, Nicolas Usunier, Alexander
  Kirillov, and Sergey Zagoruyko. 2020.
\newblock End-to-end object detection with transformers.
\newblock In \emph{European Conference on Computer Vision}, pages 213--229.
  Springer.

\bibitem[{Chang et~al.(2011)Chang, Yeh, and Miller}]{Chang2011Associating}
Tsung-Hsiang Chang, Tom Yeh, and Rob Miller. 2011.
\newblock \href {https://doi.org/10.1145/2047196.2047228} {Associating the
  visual representation of user interfaces with their internal structures and
  metadata}.
\newblock In \emph{Proceedings of the 24th Annual ACM Symposium on User
  Interface Software and Technology}, UIST '11, page 245–256, New York, NY,
  USA. Association for Computing Machinery.

\bibitem[{Chang et~al.(2010)Chang, Yeh, and Miller}]{Chang2015GUI}
Tsung-Hsiang Chang, Tom Yeh, and Robert~C. Miller. 2010.
\newblock \href {https://doi.org/10.1145/1753326.1753555} {\emph{GUI Testing
  Using Computer Vision}}, page 1535–1544. Association for Computing
  Machinery, New York, NY, USA.

\bibitem[{Chen et~al.(2020{\natexlab{a}})Chen, Chen, Xing, Xu, Zhut, Li, and
  Wang}]{Chen2020Unblind}
Jieshan Chen, Chunyang Chen, Zhenchang Xing, Xiwei Xu, Liming Zhut, Guoqiang
  Li, and Jinshui Wang. 2020{\natexlab{a}}.
\newblock Unblind your apps: Predicting natural-language labels for mobile gui
  components by deep learning.
\newblock In \emph{2020 IEEE/ACM 42nd International Conference on Software
  Engineering (ICSE)}, pages 322--334.

\bibitem[{Chen et~al.(2020{\natexlab{b}})Chen, Xie, Xing, Chen, Xu, Zhu, and
  Li}]{Chen21Object}
Jieshan Chen, Mulong Xie, Zhenchang Xing, Chunyang Chen, Xiwei Xu, Liming Zhu,
  and Guoqiang Li. 2020{\natexlab{b}}.
\newblock \href {https://doi.org/10.1145/3368089.3409691} {Object detection for
  graphical user interface: Old fashioned or deep learning or a combination?}
\newblock ESEC/FSE 2020, page 1202–1214, New York, NY, USA. Association for
  Computing Machinery.

\bibitem[{Deka et~al.(2017)Deka, Huang, Franzen, Hibschman, Afergan, Li,
  Nichols, and Kumar}]{Deka17Rico}
Biplab Deka, Zifeng Huang, Chad Franzen, Joshua Hibschman, Daniel Afergan, Yang
  Li, Jeffrey Nichols, and Ranjitha Kumar. 2017.
\newblock Rico: A mobile app dataset for building data-driven design
  applications.
\newblock In \emph{Proceedings of the 30th Annual Symposium on User Interface
  Software and Technology}, UIST '17.

\bibitem[{Deka et~al.(2016)Deka, Huang, and Kumar}]{Deka2016Erica}
Biplab Deka, Zifeng Huang, and Ranjitha Kumar. 2016.
\newblock \href {https://doi.org/10.1145/2984511.2984581} {Erica: Interaction
  mining mobile apps}.
\newblock In \emph{Proceedings of the 29th Annual Symposium on User Interface
  Software and Technology}, UIST '16, pages 767--776, New York, NY, USA. ACM.

\bibitem[{Deng et~al.(2009)Deng, Dong, Socher, Li, Li, and
  Fei-Fei}]{Deng2009ImageNet}
Jia Deng, Wei Dong, Richard Socher, Li-Jia Li, Kai Li, and Li~Fei-Fei. 2009.
\newblock \href {https://doi.org/10.1109/CVPR.2009.5206848} {Imagenet: A
  large-scale hierarchical image database}.
\newblock In \emph{2009 IEEE Conference on Computer Vision and Pattern
  Recognition}, pages 248--255.

\bibitem[{Duan et~al.(2019)Duan, Bai, Xie, Qi, Huang, and
  Tian}]{duan2019centernet}
Kaiwen Duan, Song Bai, Lingxi Xie, Honggang Qi, Qingming Huang, and Qi~Tian.
  2019.
\newblock Centernet: Keypoint triplets for object detection.
\newblock In \emph{Proceedings of the IEEE/CVF international conference on
  computer vision}, pages 6569--6578.

\bibitem[{He et~al.(2016)He, Zhang, Ren, and Sun}]{He2016Resnet}
Kaiming He, Xiangyu Zhang, Shaoqing Ren, and Jian Sun. 2016.
\newblock Deep residual learning for image recognition.
\newblock In \emph{Proceedings of the IEEE conference on computer vision and
  pattern recognition}, pages 770--778.

\bibitem[{He et~al.(2021)He, Sunkara, Zang, Xu, Liu, Wichers, Schubiner, Lee,
  and Chen}]{He2021ActionBert}
Zecheng He, Srinivas Sunkara, Xiaoxue Zang, Ying Xu, Lijuan Liu, Nevan Wichers,
  Gabriel Schubiner, Ruby Lee, and Jindong Chen. 2021.
\newblock \href {https://ojs.aaai.org/index.php/AAAI/article/view/16741}
  {Actionbert: Leveraging user actions for semantic understanding of user
  interfaces}.
\newblock \emph{Proceedings of the AAAI Conference on Artificial Intelligence},
  35(7):5931--5938.

\bibitem[{Huang et~al.(2019)Huang, Canny, and Nichols}]{Huang2019Swire}
Forrest Huang, John~F. Canny, and Jeffrey Nichols. 2019.
\newblock \href {https://doi.org/10.1145/3290605.3300334} {\emph{Swire:
  Sketch-Based User Interface Retrieval}}, page 1–10. Association for
  Computing Machinery, New York, NY, USA.

\bibitem[{Hurst et~al.(2010)Hurst, Hudson, and
  Mankoff}]{Hurst2010Auotmatically}
Amy Hurst, Scott~E. Hudson, and Jennifer Mankoff. 2010.
\newblock \href {https://doi.org/10.1145/1719970.1719973} {Automatically
  identifying targets users interact with during real world tasks}.
\newblock In \emph{Proceedings of the 15th International Conference on
  Intelligent User Interfaces}, IUI '10, page 11–20, New York, NY, USA.
  Association for Computing Machinery.

\bibitem[{Kingma and Ba(2014)}]{kingma2014adam}
Diederik~P Kingma and Jimmy Ba. 2014.
\newblock Adam: A method for stochastic optimization.
\newblock \emph{arXiv preprint arXiv:1412.6980}.

\bibitem[{Krizhevsky et~al.(2012)Krizhevsky, Sutskever, and
  Hinton}]{krizhevsky2012Alexnet}
Alex Krizhevsky, Ilya Sutskever, and Geoffrey~E Hinton. 2012.
\newblock Imagenet classification with deep convolutional neural networks.
\newblock \emph{Advances in neural information processing systems},
  25:1097--1105.

\bibitem[{Ladner(2015)}]{Ladner15Design}
Richard~E. Ladner. 2015.
\newblock \href {https://doi.org/10.1145/2723869} {Design for user
  empowerment}.
\newblock \emph{Interactions}, 22(2):24–29.

\bibitem[{Lan et~al.(2019)Lan, Chen, Goodman, Gimpel, Sharma, and
  Soricut}]{lan2019albert}
Zhenzhong Lan, Mingda Chen, Sebastian Goodman, Kevin Gimpel, Piyush Sharma, and
  Radu Soricut. 2019.
\newblock Albert: A lite bert for self-supervised learning of language
  representations.
\newblock \emph{arXiv preprint arXiv:1909.11942}.

\bibitem[{Leiva et~al.(2020)Leiva, Hota, and Oulasvirta}]{Leiva2020Enrico}
Luis~A. Leiva, Asutosh Hota, and Antti Oulasvirta. 2020.
\newblock \href {https://doi.org/10.1145/3406324.3410710} {Enrico: A dataset
  for topic modeling of mobile ui designs}.
\newblock In \emph{22nd International Conference on Human-Computer Interaction
  with Mobile Devices and Services}, MobileHCI '20, New York, NY, USA.
  Association for Computing Machinery.

\bibitem[{Li et~al.(2022)Li, Baechler, Tragut, and
  Li}]{li2022learningtodenoise}
Gang Li, Gilles Baechler, Manuel Tragut, and Yang Li. 2022.
\newblock \href {https://doi.org/10.48550/ARXIV.2201.04100} {Learning to
  denoise raw mobile ui layouts for improving datasets at scale}.

\bibitem[{Li et~al.(2021)Li, Popowski, Mitchell, and Myers}]{Li2021Screen}
Toby Jia-Jun Li, Lindsay Popowski, Tom Mitchell, and Brad~A Myers. 2021.
\newblock \href {https://doi.org/10.1145/3411764.3445049} {\emph{Screen2Vec:
  Semantic Embedding of GUI Screens and GUI Components}}. Association for
  Computing Machinery, New York, NY, USA.

\bibitem[{Li et~al.(2020{\natexlab{a}})Li, He, Zhou, Zhang, and
  Baldridge}]{li2020RicoSCA}
Yang Li, Jiacong He, Xin Zhou, Yuan Zhang, and Jason Baldridge.
  2020{\natexlab{a}}.
\newblock \href {https://doi.org/10.18653/v1/2020.acl-main.729} {Mapping
  natural language instructions to mobile {UI} action sequences}.
\newblock In \emph{Proceedings of the 58th Annual Meeting of the Association
  for Computational Linguistics}, pages 8198--8210, Online. Association for
  Computational Linguistics.

\bibitem[{Li et~al.(2020{\natexlab{b}})Li, Li, He, Zheng, Li, and
  Guan}]{Li2020Widget}
Yang Li, Gang Li, Luheng He, Jingjie Zheng, Hong Li, and Zhiwei Guan.
  2020{\natexlab{b}}.
\newblock \href {https://doi.org/10.18653/v1/2020.emnlp-main.443} {Widget
  captioning: Generating natural language description for mobile user interface
  elements}.
\newblock In \emph{Proceedings of the 2020 Conference on Empirical Methods in
  Natural Language Processing (EMNLP)}, pages 5495--5510, Online. Association
  for Computational Linguistics.

\bibitem[{Lin et~al.(2017)Lin, Doll{\'a}r, Girshick, He, Hariharan, and
  Belongie}]{lin2017fpn}
Tsung-Yi Lin, Piotr Doll{\'a}r, Ross Girshick, Kaiming He, Bharath Hariharan,
  and Serge Belongie. 2017.
\newblock Feature pyramid networks for object detection.
\newblock In \emph{Proceedings of the IEEE conference on computer vision and
  pattern recognition}, pages 2117--2125.

\bibitem[{Liu et~al.(2018)Liu, Craft, Situ, Yumer, Mech, and
  Kumar}]{Liu2018LearningDesign}
Thomas~F. Liu, Mark Craft, Jason Situ, Ersin Yumer, Radomir Mech, and Ranjitha
  Kumar. 2018.
\newblock \href {https://doi.org/10.1145/3242587.3242650} {Learning design
  semantics for mobile apps}.
\newblock In \emph{Proceedings of the 31st Annual ACM Symposium on User
  Interface Software and Technology}, UIST '18, page 569–579, New York, NY,
  USA. Association for Computing Machinery.

\bibitem[{Liu(2020)}]{Liu2020Discovering}
Zhe Liu. 2020.
\newblock \href {https://doi.org/10.1145/3324884.3418917} {Discovering ui
  display issues with visual understanding}.
\newblock In \emph{Proceedings of the 35th IEEE/ACM International Conference on
  Automated Software Engineering}, ASE '20, page 1373–1375, New York, NY,
  USA. Association for Computing Machinery.

\bibitem[{Newell et~al.(2016)Newell, Yang, and Deng}]{Newell2016hourglass}
Alejandro Newell, Kaiyu Yang, and Jia Deng. 2016.
\newblock Stacked hourglass networks for human pose estimation.
\newblock In \emph{Computer Vision -- ECCV 2016}, pages 483--499, Cham.
  Springer International Publishing.

\bibitem[{Nguyen et~al.(2018)Nguyen, Vu, Pham, and Nguyen}]{Tam2018Deep}
Tam~The Nguyen, Phong~Minh Vu, Hung~Viet Pham, and Tung~Thanh Nguyen. 2018.
\newblock \href {https://doi.org/10.1145/3183399.3183422} {Deep learning ui
  design patterns of mobile apps}.
\newblock In \emph{Proceedings of the 40th International Conference on Software
  Engineering: New Ideas and Emerging Results}, ICSE-NIER '18, page 65–68,
  New York, NY, USA. Association for Computing Machinery.

\bibitem[{Organization(2021)}]{WHO21Blindness}
World~Health Organization. 2021.
\newblock \href
  {https://www.who.int/news-room/fact-sheets/detail/blindness-and-visual-impairment}
  {Blindness and vision impairment}.

\bibitem[{Peter~Ackland and Bourne(2017)}]{Ackland17World}
Serge~Resnikoff Peter~Ackland and Rupert Bourne. 2017.
\newblock World blindness and visual impairment: despite many successes, the
  problem is growing. community eye health.
\newblock \emph{Community Eye Health}, 30(100):71--73.

\bibitem[{Ren et~al.(2015)Ren, He, Girshick, and Sun}]{ren2015frcnn}
Shaoqing Ren, Kaiming He, Ross Girshick, and Jian Sun. 2015.
\newblock Faster r-cnn: Towards real-time object detection with region proposal
  networks.
\newblock \emph{Advances in neural information processing systems}, 28:91--99.

\bibitem[{Ross et~al.(2017)Ross, Zhang, Fogarty, and
  Wobbrock}]{Ross2017Epidemiology}
Anne~Spencer Ross, Xiaoyi Zhang, James Fogarty, and Jacob~O. Wobbrock. 2017.
\newblock \href {https://doi.org/10.1145/3132525.3132547} {Epidemiology as a
  framework for large-scale mobile application accessibility assessment}.
\newblock In \emph{Proceedings of the 19th International ACM SIGACCESS
  Conference on Computers and Accessibility}, ASSETS '17, page 2–11, New
  York, NY, USA. Association for Computing Machinery.

\bibitem[{Ross et~al.(2020)Ross, Zhang, Fogarty, and
  Wobbrock}]{Ross2020Epidemiology}
Anne~Spencer Ross, Xiaoyi Zhang, James Fogarty, and Jacob~O. Wobbrock. 2020.
\newblock \href {https://doi.org/10.1145/3348797} {An epidemiology-inspired
  large-scale analysis of android app accessibility}.
\newblock \emph{ACM Trans. Access. Comput.}, 13(1).

\bibitem[{Szegedy et~al.(2017)Szegedy, Ioffe, Vanhoucke, and
  Alemi}]{szegedy2017inception}
Christian Szegedy, Sergey Ioffe, Vincent Vanhoucke, and Alexander~A Alemi.
  2017.
\newblock Inception-v4, inception-resnet and the impact of residual connections
  on learning.
\newblock In \emph{Thirty-first AAAI conference on artificial intelligence}.

\bibitem[{Tan and Le(2019)}]{tan2019efficientnet}
Mingxing Tan and Quoc Le. 2019.
\newblock Efficientnet: Rethinking model scaling for convolutional neural
  networks.
\newblock In \emph{International Conference on Machine Learning}, pages
  6105--6114. PMLR.

\bibitem[{Vaswani et~al.(2017)Vaswani, Shazeer, Parmar, Uszkoreit, Jones,
  Gomez, Kaiser, and Polosukhin}]{Vaswani2017Attention}
Ashish Vaswani, Noam Shazeer, Niki Parmar, Jakob Uszkoreit, Llion Jones,
  Aidan~N Gomez, \L~ukasz Kaiser, and Illia Polosukhin. 2017.
\newblock \href
  {https://proceedings.neurips.cc/paper/2017/file/3f5ee243547dee91fbd053c1c4a845aa-Paper.pdf}
  {Attention is all you need}.
\newblock In \emph{Advances in Neural Information Processing Systems},
  volume~30. Curran Associates, Inc.

\bibitem[{XCode(2022)}]{XCode2022iOS}
XCode. 2022.
\newblock \href {https://developer.apple.com/xcode/} {Xcode}.

\bibitem[{Yeh et~al.(2009)Yeh, Chang, and Miller}]{Yeh2009Sikuli}
Tom Yeh, Tsung-Hsiang Chang, and Robert~C. Miller. 2009.
\newblock \href {https://doi.org/10.1145/1622176.1622213} {Sikuli: Using gui
  screenshots for search and automation}.
\newblock In \emph{Proceedings of the 22nd Annual ACM Symposium on User
  Interface Software and Technology}, UIST '09, page 183–192, New York, NY,
  USA. Association for Computing Machinery.

\bibitem[{Zhang et~al.(2021)Zhang, de~Greef, Swearngin, White, Murray, Yu,
  Shan, Nichols, Wu, Fleizach, Everitt, and Bigham}]{Zhang2021Screen}
Xiaoyi Zhang, Lilian de~Greef, Amanda Swearngin, Samuel White, Kyle Murray,
  Lisa Yu, Qi~Shan, Jeffrey Nichols, Jason Wu, Chris Fleizach, Aaron Everitt,
  and Jeffrey~P Bigham. 2021.
\newblock \href {https://doi.org/10.1145/3411764.3445186} {Screen recognition:
  Creating accessibility metadata for mobile applications from pixels}.
\newblock In \emph{Proceedings of the 2021 CHI Conference on Human Factors in
  Computing Systems}, CHI '21, New York, NY, USA. Association for Computing
  Machinery.

\bibitem[{Zhou et~al.(2019)Zhou, Wang, and
  Kr{\"a}henb{\"u}hl}]{zhou2019centernet}
Xingyi Zhou, Dequan Wang, and Philipp Kr{\"a}henb{\"u}hl. 2019.
\newblock Objects as points.
\newblock \emph{arXiv preprint arXiv:1904.07850}.

\bibitem[{Zhu et~al.(2020)Zhu, Su, Lu, Li, Wang, and Dai}]{zhu2020deformable}
Xizhou Zhu, Weijie Su, Lewei Lu, Bin Li, Xiaogang Wang, and Jifeng Dai. 2020.
\newblock Deformable detr: Deformable transformers for end-to-end object
  detection.
\newblock \emph{arXiv preprint arXiv:2010.04159}.

\end{thebibliography}

\appendix

\section{Appendix: Data distribution and experiment results}
\label{sec:appendix-data}
This section contains more details on the datasets. Section \ref{sec:semantics-definitions} contains the definitions used for attributing  the detailed data distribution for each of the tasks. Table \ref{table:shape-stats} contains the data distribution for the 76 \emph{icon shape} classes. Table \ref{table:semantics-stats} contains the shape classes and their semantic classification along with counts for each class and table \ref{table:grouping-stats} contains the data distribution for the \emph{label association} classes.

In addition Table \ref{table:od-expanded-results} contains results for object detection models based on Faster R-CNN on the 3 tasks discussed in the paper.

\subsection{Semantic class definitions}
\label{sec:semantics-definitions}
As mentioned in section \ref{sec:icon-semantics} the shape icons are further sub-divided into various categories based on their functionality. The definitions of the various semantic types are given below. Each semantic icon name is prefixed by the corresponding shape name. We exclude the OTHER category for each icon shape as it is used to capture all other functionalities not covered by the mentioned semantics.
\begin{itemize}[noitemsep,topsep=0pt,parsep=0pt,partopsep=0pt]
    \item ICON\_X:CLOSE Close windows or tabs or exit a window.
    \item ICON\_X:DELETE TEXT Delete entries, items, text, suggestions etc.
    \item ICON\_X:MULTIPLY Mathematical operation of multiplication.
    \item ICON\_ARROW\_UPWARD:CAPITALIZE Caps Lock icon to toggle upper case and lower case letters in the keyboard.
    \item ICON\_MAGNIFYING\_GLASS:SEARCH Search in the current app or website.
    \item ICON\_MAGNIFYING\_GLASS:ZOOM IN Zoom-in to a picture, document etc.
    \item ICON\_MAGNIFYING\_GLASS:ZOOM OUT Zoom-out of a picture, document etc.
    \item ICON\_UNDO:REPLY Reply to a message, mail etc.
    \item ICON\_UNDO:UNDO Undo the previous action.
    \item ICON\_UNDO:BACK Go back to the previous screen or state.
    \item ICON\_REDO:SHARE Share this item.
    \item ICON\_REDO:REDO Redo the previous action.
    \item ICON\_THREE\_BARS:MENU Icon to display menu options.
    \item ICON\_PHONE:CALL Start a phone call.
    \item ICON\_PHONE:CHAT APP Icon for a chat app.
    \item ICON\_PHONE:PHONE APP Open the phone app.
    \item ICON\_PHONE:END CALL End a phone or video call.
    \item ICON\_PLAY:PLAY Playing video, audio, games, etc.
    \item ICON\_PLAY:PLAY STORE Icon for the Google Play Store.
    \item ICON\_PLAY:YOUTUBE Icon for the YouTube app.
    \item ICON\_CHAT:CHAT Send a message to someone or view comments.
    \item ICON\_CHAT:WHATSAPP Icon for the WhatsApp app.
    \item ICON\_CHAT:FACEBOOK MESSENGER Icon for the Facebook Messenger app.
    \item ICON\_TAKE\_PHOTO:INSTAGRAM Icon for the Instagram app.
    \item ICON\_THREE\_DOTS:MORE For “more” options, contents, etc. It could also refer to menu.
    \item ICON\_PLUS:ADD ITEM Add a new item to an existing list.
    \item ICON\_PLUS:EXPAND Expand a UI element to show more details.
\end{itemize}

\noindent\hrulefill\par
\begin{table*}[t!]
\noindent\makebox[0.5\textwidth][c]{%
\begin{minipage}[b]{0.45\textwidth}
\begin{tabular}{l r}
\toprule
\textbf{Shape Class} & \textbf{Count} \\
\toprule
ICON V BACKWARD & 46,431\\ 
ICON NAV BAR CIRCLE & 41,551\\ 
ICON NAV BAR RECT & 41,449\\ 
ICON STAR & 17,890\\ 
ICON THREE DOTS & 15,194\\ 
ICON V FORWARD & 14,131\\ 
ICON ARROW BACKWARD & 13,767\\ 
ICON THREE BARS & 13,659\\ 
ICON X & 11,058\\ 
ICON MAGNIFYING GLASS & 10,911\\ 
ICON PLUS & 9,971\\ 
ICON PLAY & 7,576\\ 
ICON V DOWNWARD & 7,447\\ 
ICON PERSON & 6,648\\ 
ICON CHECK & 6,583\\ 
ICON HEART & 6,274\\ 
ICON CHAT & 5,483\\ 
ICON SETTINGS & 4,909\\ 
ICON SHARE & 4,871\\ 
ICON ARROW FORWARD & 3,463\\ 
ICON LOCATION & 3,398\\ 
ICON INFO & 3,287\\ 
ICON HOME & 3,172\\ 
ICON TIME & 3,123\\ 
ICON REFRESH & 2,987\\
ICON CLOUD & 2,436\\ 
ICON EDIT & 2,280\\ 
ICON QUESTION & 2,263\\ 
ICON TAKE PHOTO & 2,110\\ 
ICON SHOPPING CART & 1,900\\ 
ICON CALENDAR & 1,851\\ 
ICON NOTIFICATIONS & 1,817\\
ICON CLOUD & 2,436\\ 
ICON EDIT & 2,280\\ 
ICON QUESTION & 2,263\\ 
ICON TAKE PHOTO & 2,110\\ 
ICON SHOPPING CART & 1,900\\ 
ICON CALENDAR & 1,851\\ 
ICON NOTIFICATIONS & 1,817\\ 
ICON FACEBOOK & 1,700\\ 
ICON ENVELOPE & 1,659\\ 
ICON PEOPLE & 1,658\\ \toprule
\end{tabular}
\end{minipage}}
\noindent\makebox[0.5\textwidth][c]{%
\begin{minipage}[b]{0.45\textwidth}
\begin{tabular}{l r}
\toprule
\textbf{Shape Class} & \textbf{Count} \\
\toprule
ICON LOCK & 1,622\\ 
ICON GALLERY & 1,535\\ 
ICON CALL & 1,488\\ 
ICON V UPWARD & 1,392\\ 
ICON VOLUME STATE & 1,359\\
ICON LIST & 1,346\\ 
ICON DOWNLOAD & 1,344\\ 
ICON THUMBS UP & 1,335\\ 
ICON SUN & 1,327\\ 
ICON ARROW DOWNWARD & 1,317\\ 
ICON LAUNCH APPS & 1,136\\ 
ICON ARROW UPWARD & 1,094\\ 
ICON MIC & 1,016\\ 
ICON HAPPY FACE & 955\\ 
ICON PAUSE & 864\\ 
ICON TWITTER & 860\\ 
ICON SHOPPING BAG & 776\\ 
ICON MOON & 719\\ 
ICON SEND & 711\\ 
ICON COMPASS & 691\\ 
ICON DELETE & 665\\ 
ICON REDO & 546\\ 
ICON VIDEOCAM & 521\\ 
ICON HISTORY & 447\\ 
ICON UNDO & 441\\ 
ICON HEADSET & 412\\ 
ICON THUMBS DOWN & 382\\ 
ICON EXPAND & 356\\ 
ICON GOOGLE & 334\\ 
ICON UPLOAD & 328\\ 
ICON SAD FACE & 239\\ 
ICON STOP & 204\\ 
ICON CAST & 150\\ 
ICON PAPERCLIP & 139\\ 
ICON VOLUME MUTE & 77\\ 
ICON END CALL & 65\\ 
ICON VOLUME DOWN & 21\\ 
ICON CONTRACT & 19\\ 
ICON VOLUME UP & 14\\ 
ICON MIC MUTE & 13\\ 
ICON ASSISTANT & 4 \\ \hline
\textbf{TOTAL} & \textbf{353,171} \\ \toprule
\end{tabular}
\end{minipage}}
\caption{Number of instances for each icon class for the \emph{Icon Shape} annotations.}
\label{table:shape-stats}
\end{table*}

\begin{table*}[t!]
\begin{center}
\begin{tabular}{l l r} 
\toprule
\textbf{Shape Class} & \textbf{Semantic Class} & \textbf{Count} \\
\toprule
\multirow{4}{*}{ICON X} & CLOSE & 8,899 \\ 
& DELETE TEXT & 1,163 \\ 
& MULTIPLY & 50 \\ 
& OTHER & 840 \\ \hline
\multirow{2}{*}{ICON ARROW UPWARD} & CAPITALIZE & 154 \\ 
& OTHER & 915 \\ \hline
\multirow{4}{*}{ICON MAGNIFYING GLASS} & SEARCH & 10,243 \\ 
& ZOOM IN & 142 \\ 
& ZOOM OUT & 92 \\ 
& OTHER & 331 \\ \hline
\multirow{4}{*}{ICON UNDO} & REPLY & 113 \\ 
& UNDO & 109 \\ 
& BACK & 101 \\ 
& OTHER & 115 \\ \hline
\multirow{3}{*}{ICON REDO} & SHARE & 354 \\
& REDO & 63 \\ 
& OTHER & 117 \\ \hline
\multirow{2}{*}{ICON THREE BARS} & MENU & 11,929 \\ 
& OTHER & 1,329 \\ \hline
\multirow{5}{*}{ICON PHONE} & CALL & 729 \\ 
& CHAT APP & 530 \\ 
& PHONE APP & 89 \\ 
& END CALL & 12 \\ 
& OTHER & 571 \\ \hline
\multirow{4}{*}{ICON CHAT} & CHAT APP & 530 \\ 
& WHATSAPP & 120 \\ 
& FACEBOOK MESSENGER & 61 \\ 
& OTHER & 4,663 \\ \hline
\multirow{2}{*}{ICON CAMERA} & INSTAGRAM & 192 \\ 
& OTHER & 1,884 \\ \hline
\multirow{4}{*}{ICON PLAY} & PLAY & 4,452 \\ 
& PLAY STORE & 782 \\ 
& YOUTUBE & 356 \\ 
& OTHER & 1,846 \\ \hline
\multirow{2}{*}{ICON THREE DOTS} & MORE & 14,285 \\ 
& OTHER & 785 \\ \hline
\multirow{4}{*}{ICON PLUS} & ADD ITEM & 7,170 \\ 
& EXPAND & 396 \\ 
& OTHER & 2,244 \\ \hline
\textbf{TOTAL} & \centering \textbf{-} & \textbf{78,756} \\
\toprule
\end{tabular}
\end{center}
\caption{Classification of \emph{Icon Shape} classes into semantic classes.}
\label{table:semantics-stats}
\end{table*}

\begin{table*}[t]
\begin{center}
\begin{tabular}{l r r r p{0cm}}
\toprule
\textbf{Class name} & \textbf{Count} & \centering \textbf{\# with text labels} & \centering \textbf{\% with text labels} & \\ 
\toprule
Icon & 252,342 & \centering 57,716 & \centering 22.87\% & \\ 
Text Field & 16,131 & \centering 3,292 & \centering 20.41\% & \\ 
Check Box & 5,958 & \centering 3,723 &  \centering 62.49\% & \\
Radio Button & 2,558 & \centering 1,659 &  \centering 64.86\% & \\
\hline
\textbf{Total} & \textbf{276,989} & \centering \textbf{66,390} & \centering \textbf{23.96\%} & \\
\toprule
\end{tabular}
\end{center}
\caption{\emph{Label Association} task statistics indicating the overall counts of the different classes and the frequency with which text labels are associated with them.}
\label{table:grouping-stats}
\end{table*}

\begin{table*}[t]
\centering
\begin{tabular}{l c c c c c c} 
\toprule
 &  \multicolumn{2}{c}{Icon Shape} & \multicolumn{2}{c}{Icon Semantics} & \multicolumn{2}{c}{Label Association} \\
\toprule
Backbone & mAP & mAP@0.5IOU & mAP & mAP@0.5IOU &mAP & mAP@0.5IOU  \\
\hline
ResNet-101 & 32.07 & 65.72 & \textbf{25.33} & \textbf{53.59} & 34.70 & 73.37 \\
Inception ResNet & 31.61 & 70.14 & 25.17 & 53.36 & 36.63 & 75.19 \\
ResNet-101 with FPN & \textbf{34.60} & \textbf{70.24} & 25.34 & 53.17 & \textbf{36.90} & \textbf{75.75} \\
\hline
\end{tabular}
\caption{Object Detection model performance for Faster R-CNN based models with different backbone networks. The numbers in bold indicate the backbone with the best mAP@0.5IOU for the task.}
\label{table:od-expanded-results}
\end{table*}
\end{document}